\documentclass[sigconf]{acmart}

\AtBeginDocument{%
  }

\copyrightyear{2025}
\acmYear{2025}
\setcopyright{acmlicensed}\acmConference[FAccT '25]{The 2025 ACM Conference on Fairness, Accountability, and Transparency}{June 23--26, 2025}{Athens, Greece}
\acmBooktitle{The 2025 ACM Conference on Fairness, Accountability, and Transparency (FAccT '25), June 23--26, 2025, Athens, Greece}
\acmDOI{10.1145/3715275.3732185}
\acmISBN{979-8-4007-1482-5/2025/06}

\usepackage{xcolor}

\usepackage{subcaption}
\usepackage{graphicx}
\usepackage{adjustbox}
\usepackage{array}
\usepackage{enumitem}
\usepackage{multirow}
\usepackage{float}
\usepackage{placeins}
\begin{document}

\title{Privacy of Groups in Dense Street Imagery}

\author{Matt Franchi}
\authornote{The authors contributed equally to this research.}
\email{mattfranchi@cs.cornell.edu}
\orcid{}
\affiliation{%
  \institution{Cornell University, Cornell Tech}
  \city{New York}
  \state{New York}
  \country{USA}
}

\author{Hauke Sandhaus}
\email{hgs52@cornell.edu}
\orcid{0000-0002-4169-0197}
\authornotemark[1]
\affiliation{
    \institution{Cornell University, Cornell Tech}
    \city{New York}
    \state{New York}
    \country{USA}
}

\author{Madiha Zahrah Choksi}
\email{mc2376@cornell.edu}
\orcid{}
\affiliation{
    \institution{Cornell University, Cornell Tech}
    \city{New York}
    \state{New York}
    \country{USA}
}

\author{Severin Engelmann}
\email{severin.engelmann@cornell.edu}
\orcid{}
\affiliation{
    \institution{Cornell University, Cornell Tech}
    \city{New York}
    \state{New York}
    \country{USA}
}

\author{Wendy Ju}
\email{wendyju@cornell.edu}
\orcid{}
\affiliation{
    \institution{Jacobs Technion-Cornell Institute, Cornell Tech}
    \city{New York}
    \state{New York}
    \country{USA}
}

\author{Helen Nissenbaum}
\email{hn288@cornell.edu}
\orcid{}
\affiliation{
    \institution{Cornell University, Cornell Tech}
    \city{New York}
    \state{New York}
    \country{USA}}

\renewcommand{\shortauthors}{Franchi and Sandhaus et al.}

\newcommand{\numDashcamImages}{25,232,608\space}

\renewcommand{\sectionautorefname}{Section}
\renewcommand{\subsectionautorefname}{Section}
\renewcommand{\subsubsectionautorefname}{Section}

\begin{abstract}
Spatially and temporally dense street imagery (DSI) datasets have grown unbounded. In 2024, individual companies possessed around 3 \textit{trillion} unique images of public streets. DSI data streams are only set to grow as companies like Lyft and Waymo use DSI to train autonomous vehicle algorithms and analyze collisions. Academic researchers leverage DSI to explore novel approaches to urban analysis. Despite good-faith efforts by DSI providers to protect individual privacy through blurring faces and license plates, these measures fail to address broader privacy concerns. In this work, we find that increased data density and advancements in artificial intelligence enable harmful group membership inferences from supposedly anonymized data. We perform a penetration test to demonstrate how easily sensitive group affiliations can be inferred from obfuscated pedestrians in \numDashcamImages dashcam images taken in New York City. We develop a typology of identifiable groups within DSI and analyze privacy implications through the lens of contextual integrity. Finally, we discuss actionable recommendations for researchers working with data from DSI providers.   
\end{abstract}



\begin{teaserfigure}
  \centering
  \includegraphics[width=0.9\textwidth]{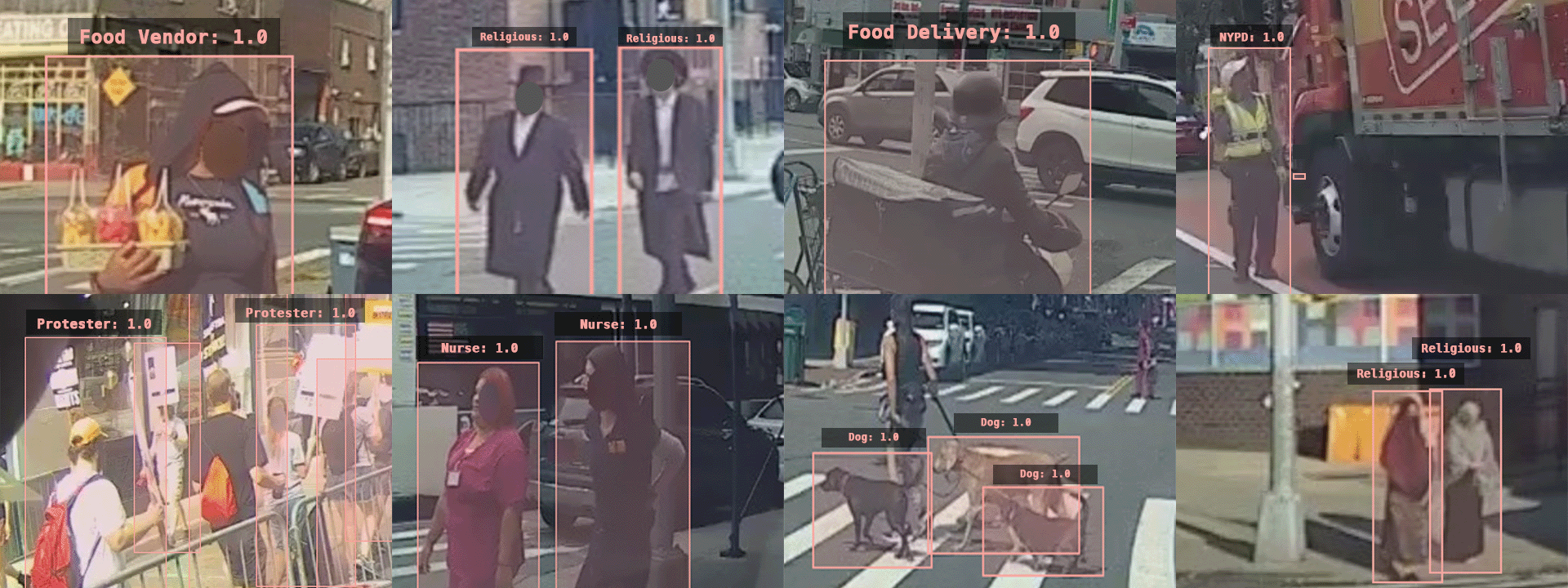}
  \caption{AI-inferred group membership in a dataset of more than 25 million facially de-identified dashcam images from NYC in 2023. A project website is available at \href{https://dsi.tech.cornell.edu/}{dsi.tech.cornell.edu}.}
  \Description{Inferred members of groups in a set of 25 million dashcam images from NYC in 2023.}
  \label{fig:teaser}
\end{teaserfigure}

\maketitle

\section{Introduction}
Dense Street Imagery (DSI) represents a major breakthrough in vehicle hardware, imaging technology, and networking, enabling dynamic, real-time depictions of locations worldwide. Unlike traditional static snapshots, such as those provided by Google Street View, DSI capitalizes on innovations in \textit{temporal} density—achieved through networked dashcams \cite{franchi_towards_2024} and advanced driver-assist systems \cite{hind_automotive_2024} to deliver fresh, continuous imagery at an unparalleled frequency. Researchers have found many beneficial applications of DSI; these include tracking transient weather events~\cite{tran_forest-fire_2022, franchi_bayesian_2025}, documenting sidewalk scaffolding~\cite{shapira_dorin_fingerprinting_2024}, analyzing vehicle placement patterns~\cite{franchi_detecting_2023}, and assessing dynamic streetconditions~\cite{dadashova_detecting_2021, franchi_robotability_2025}. By transforming mobile cameras into a distributed sensing network, DSI offers researchers new ways to understand the rapidly changing physical and social landscapes of urban environments in ways that are more flexible and adaptable than possible with location-fixed sensing systems.

The shift from periodic to near-continuous image capture has made it possible to monitor people at much shorter intervals; cycles that were once difficult or even impossible to observe. Unlike Google Street View (GSV), which faced several privacy controversies in the early 2000s, DSI has so far avoided similar public scrutiny. In the 2014 Canadian case Pia Grillo v. Google, the plaintiff sued Google for invading her privacy after Google Street View (GSV) published an image of her outside her home in Quebec~\cite{gallo_privacy_2020, burdon_google_2014, kawaguchi_what_2012}. Although GSV had blurred her face, it failed to obscure her license plate and home address, and the photo revealed part of her upper body. The court found that these visible details could allow others to identify her, despite the facial blurring. The case highlights how peripheral information such as license plates and addresses can undermine anonymity, even when facial features are obscured. 

Today, an expansive infrastructure has emerged in which companies like Mobileye manage vast datasets (reportedly 200 petabytes of data)~\cite{mobileye_mobileyes_2022}, comprising over 3 trillion images from cameras mounted on consumer vehicles~\cite{hawkins_lyft_2024}). Lyft, a leading ride-sharing company, also recently gained access to \textit{200 million miles} of driving imagery~\cite{hawkins_lyft_2024}. For reference, Google reported its Street View had 220 billion images and 10 million miles of footage in 2022. DSI has, without much notice, assembled a data moat more than ten times the size. Recent research by~\citet{sandhaus2024crashdata} reveals that autonomous vehicle companies possess vast amounts of DSI data; while they are reluctant to share it openly, they have sophisticated internal methods to remotely retrieve data from their fleets. How can individuals escape the surveillance potential inherent to DSI? Increasingly, it seems that to opt out, one must opt out of public space~\cite{green_smart_2019}. Privacy defenses for sensitive objects in DSI include the blurring of faces, license plates, and other user-requested objects. Such practices have precedents in earlier technologies like Google Street View~\cite{anguelov_google_2010}. Google states: ``We have developed cutting-edge face and license plate blurring technology that is designed to blur identifiable faces and license plates within Google-contributed imagery in Street View''~\cite{google_google-contributed_2025}; this seems to be the commercially-standard protection standard~\cite{frome_large-scale_2009}\footnote{We note that at the time of writing, Nexar, the provider of our experimental dataset, goes beyond the popular commercial standard by blurring entire pedestrian figures instead of just faces (see \autoref{fig:nexar-obfuscation-present} for an example). However, this practice reduces the visual fidelity and utility of certain pedestrian-dense images, presenting an open problem.}. Privacy-preserving mechanisms for static street view technologies, including blurring faces, license plates, and other user-requested objects, are inadequate and notably blunt~\cite{burdon_object_2024}, and obfuscation failure modes are noted and exist~\cite{burdon_object_2024} -- see \autoref{sec:C-obfuscation-taxonomy} for details on specific failure modes. The inevitable increase in \textit{temporal} density further undermines these established mechanisms~\cite{franchi_towards_2024}. Ultimately, in DSI, objects of interest are \textit{traceable}. Additionally, artificial intelligence (AI)’s inferential capabilities make it possible to generate detailed information without direct collection, for example, using vision models to analyze visual data and identify individuals’ clothing types, styles, and accessories in public spaces in near real-time~\cite{cheng2021fashion, bossard2013apparel}. This allows identification of group affiliations (e.g., demonstrators, religious congregations, professionals) based on attire and accessories. Even without explicit group markers, physical proximity to others displaying group affiliation can signal group association~\cite{pyrgelis2017knock}. In public spaces, computer vision models can infer protected attributes like gender through pose estimation~\cite{kastaniotis2013gait} and disabilities using proxies like wheelchairs~\cite{rosero2018intelligent}. Moreover, cross analyses with mobility data or activity data, such as pings of individuals' cell phone location across time, further motivate privacy risks in DSI. A landmark study analyzing mobility data for 1.5 million individuals over 15 months revealed that just four spatiotemporal points were sufficient to uniquely identify 95\% of individuals in the dataset~\cite{de_montjoye_unique_2013}. These results have been replicated in subsequent studies (e.g.,~\cite{de_montjoye_unique_2015}) and highlight the significant surveillance potential of DSI. Why? Data of this nature is highly sensitive due to the uniqueness of human behavior~\cite{green_smart_2019}. DSI — with its visual dimension — reveals a heightened sensitivity, enabling AI to make inferences about appearance and behaviors.

This work pioneers the exploration of group identifiability in public street imagery, leveraging a real-world dataset of \numDashcamImages unique dashcam images captured in New York City (visualized in \autoref{fig:coverage}) provided for research evaluation by Nexar, Inc., a dashcam manufacturer and smart-mapping startup. We note that this work does \textit{not} address the potential identifiability of individuals. Our study intersects penetration testing, group privacy, and contextual integrity to investigate DSI's implications for society. We begin with a penetration test of a real-world DSI dataset to demonstrate how face-deidentified imagery can be circumvented with ease, revealing artifacts that can lead to group privacy harms. The results of our penetration test motivate our downstream research questions. 

Next, we present related work, including an overview of DSI-producing technologies, relevant privacy theory, and examples of inferences produced by computer vision models. We then demarcate information flows within DSI, using the framework of contextual integrity. Finally, we discuss and synthesize findings, document harms, and offer recommendations for DSI data sharing and use within academia.

\section{Pentesting a dataset of DSI for risks of privacy harms}
We begin by demonstrating how efficiently DSI can be integrated into an application capable of inferring membership in a \textit{group}, all while preserving individual privacy through de-identification. How do we define `de-identification'? In the context of DSI, de-identification refers to the visual blocking of elements considered sensitive, such as blurring all faces in a dataset. DSI data providers disproportionately rely on \textit{facial blurring} as a privacy measure because of its simplicity and extendability across diverse contexts~\cite{burdon_object_2024}. Building on this, we add that individual, pedestrian-level obfuscation is not enough privacy protection. We examine this assertion through an approach borrowed from computer security literature, called \textit{penetration testing}, or `pentesting'~\cite{bishop_about_2007}. Penetration testing typically involves breaching a system to assess the difficulty of doing so. However, Bishop~\cite{bishop_about_2007} emphasizes that it should also include a thorough analysis of threats and potential attackers, an approach that aligns with our later examination of DSI information flows through the lens of contextual integrity. Rather than assuming the absence of privacy harms, our work actively identifies them~\cite{bishop_about_2007}.

\subsection{Adversarial Methodology}
In our penetration test, we define an adversary as an `authority,' such as law enforcement or government officials. Our experiments utilize a comprehensive dataset of \numDashcamImages images collected across New York City (detailed sampling methodology in \autoref{sec:B_nitty-gritty}). From an adversary's perspective, we identify several methods using DSI imagery—enabled by recent advances in artificial intelligence—that could potentially harm a group or its members. These methods are outlined in \autoref{tab:adversarial-methodology} and intended to be illustrative rather than exhaustive.

\begin{table*}[h]
    \centering
    {
\begin{tabular}{lp{7cm}p{6.5cm}}
    \toprule 
    \textbf{Method} & \textbf{Description} & \textbf{Example} \\ 
    \hline
    Zero-shot & Inferring images with a foundation model or machine learning model, and treating its output as ground truth, without any task-specific labeled data. & Using a web-trained model to identify group instances, e.g., by describing visual characteristics like "umbrellas" or 
    "LED signs." \\ 
    Supervised & Human generation of training and validation labels, training a model on these annotations, and treating its output as ground truth. & Crowdworkers annotate images with labels like "protesters" or "food vendors," train a model on these labels, and then classify new images. \\ 
    Unsupervised & Generating image embeddings from a model, clustering them, and manually identifying labels for the various clusters. & Cluster image embeddings, label visual clusters (e.g., "colorful trucks"), and apply these labels to all images in the cluster. \\
    & & \\
    \hline 
    Known-event matching & Retrieving images occurring at the same location and time as a known event, and using the image's contents as additional context for an adversarial task. & Manually matching images to a known protest time and location to interpret the scenes and identify groups. \\ 
    Geofencing & Retrieving images within a geographic region of interest, and using the contents of those images as additional context for an adversarial task. & Focusing on a city block known for street vendors, labeling initial images to define key features, and then applying those features to identify similar vendors in the area. \\
    \bottomrule
\end{tabular}
}

    \caption{DSI group identification and retrieval methods. We perform experiments based on zero-shot methods.}
    \label{tab:adversarial-methodology}
\end{table*}

\paragraph{Experiment 1}The first experiment starts with the \textit{zero-shot} method described in \autoref{tab:adversarial-methodology}. To source training data, we conducted a zero-shot image classification task on 500,000 randomly sampled images, leveraging vision-language models (VLMs) \cite{pan_zero-shot_2024, wang_zero-shot_2024}. Specifically, we used the VLM Cambrian-13B \cite{tong_cambrian-1_2024} to answer the prompt: "\textit{Is there a food truck in this image?}", receiving yes or no answers. Next, we manually validated the classified positives through human annotation\footnote{This task was carried out by a team of human annotators, including two authors of this paper, both with extensive experience observing the cultural norms and street activity of New York City.}. We also report standard model performance metrics in \autoref{sec:cambrian-model-validation}. Finally, we trained a series of lightweight YOLOv11 (\cite{jocher_ultralytics_2023}) object detection models and selected the most performant. These models, capable of real-time and distributed inference, were chosen to illustrate the ease with which imagery can be transformed into spatiotemporal distributions of inferred group members. Lastly, we estimated the spatiotemporal distribution of each group in the entire dataset of \numDashcamImages images by running inferences on each image with the trained YOLO model. We provide more information on the training of the YOLO model in \autoref{sec:yolo-model-training}. 

\paragraph{Experiment 2}The second experiment more directly encompasses the \textit{zero-shot} method from \autoref{tab:adversarial-methodology}. Similarly, we run a zero-shot image classification task on 500,000 randomly sampled images using the same VLM (Cambrian-13B). For this task, we asked Cambrian: ``\textit{Is there a bike rider with a box on their back in this image?}''\footnote{We experimented with several different prompts on a small sample of randomly sampled images and found that Cambrian has little predictive power on domain-specific terms like 'food delivery worker' or 'Uber Eats driver'. Consequently, we prompted for the flagship equipment that food delivery workers wear while biking around the city: food storage boxes strapped onto the back of a bike.}. Then, we took the Cambrian model output as ground-truth and created a detection heatmap (see \autoref{fig:food-delivery-heatmap}). We draw parallels between this approach and the biased, partially inaccurate machine learning models that have been deployed in algorithmic policing endeavors \cite{lum_predict_2016, bennett_moses_algorithmic_2018, robertson_surveil_2020}. In the following, we provide information on mobile food vending and food delivery in New York City.

\begin{figure}[h]
    \includegraphics[width=0.48\textwidth,height=0.5\textheight,keepaspectratio]{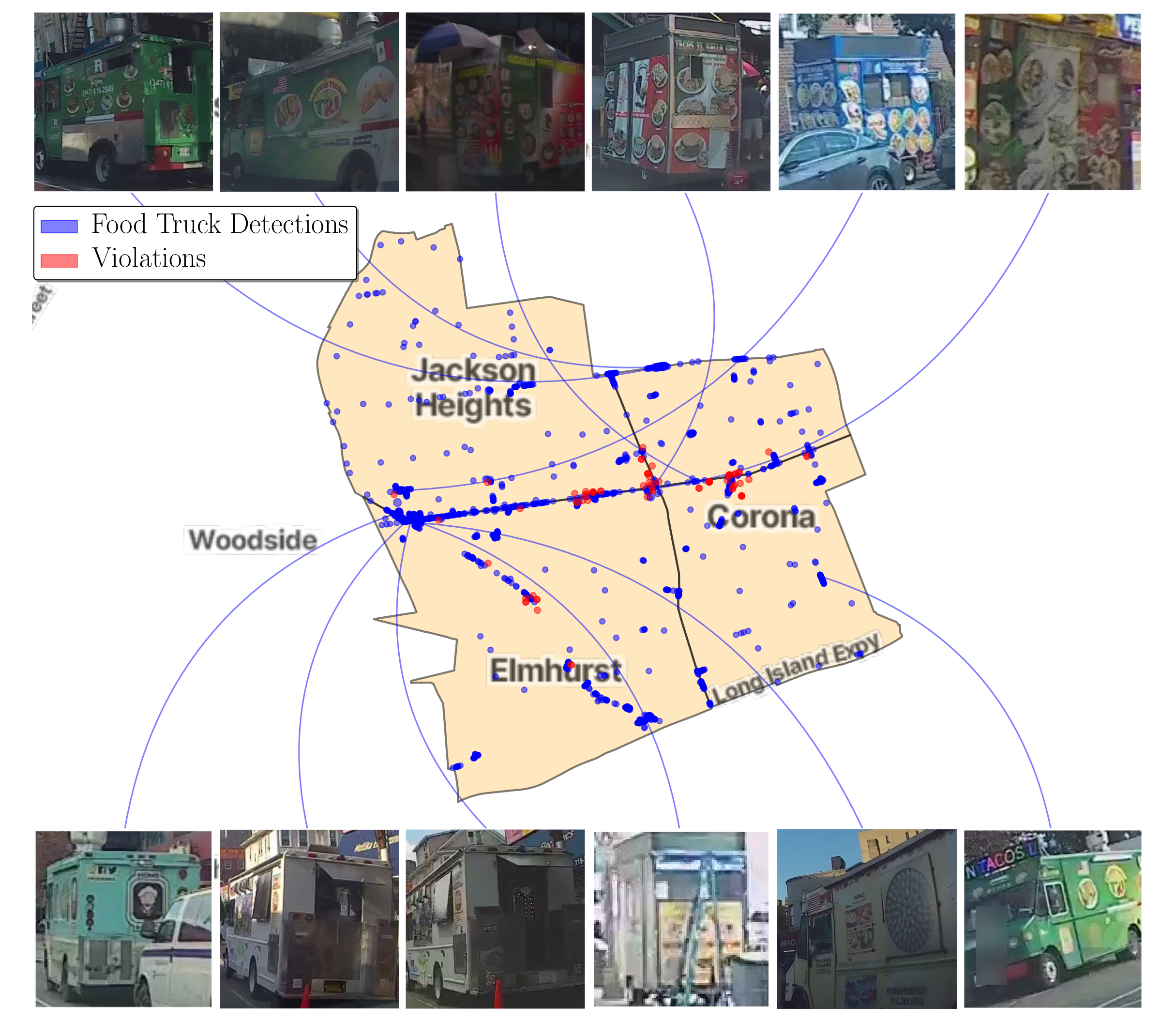}
    \caption{A map showing reported vending violations against food truck detections in Jackson Heights, Queens. For higher precision, we choose a confidence threshold of 0.7, which yields a precision of 0.90 and a recall of 0.50 on the test set.}
    \label{fig:food-truck-map}
    \Description{A map of Jackson Heights, Queens showing food truck detections as blue points and vending violations as red triangles. The map reveals significant spatial overlap between detected food trucks and reported vending violations, demonstrating how DSI surveillance could be used to target street vendors.}
\end{figure}

\begin{figure}[h!]
    \includegraphics[width=0.48\textwidth]{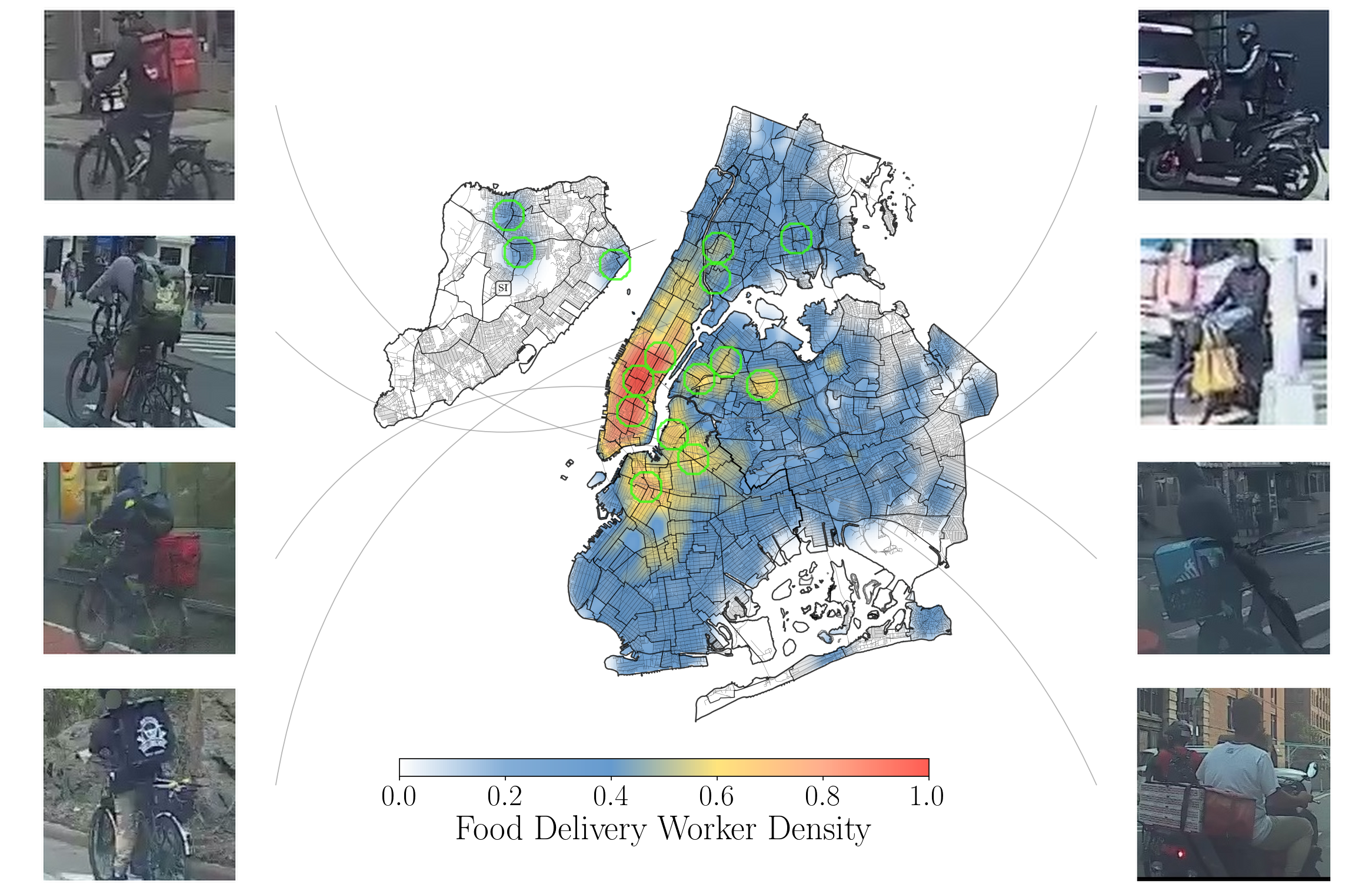}
    \caption{Using zero-shot image retrieval, we queried Cambrian for the prompt ``Is there a bike rider with a box on their back in this image?''. An authority may readily use this to create a strategic map for deployment zones optimal for monitoring food delivery worker hotspots, as depicted. The computed hotspots correspond to the average 'lunch rush' period (10AM-2PM) and can be easily computed over each day within our dataset. From a ground-truth annotation of 500 random positive detections, we estimate precision at 0.70.}
    \label{fig:food-delivery-heatmap}
    \Description{A map of New York City showing food delivery worker hotspots detected through AI analysis of dashcam imagery. The map includes several inset images showing delivery workers on bicycles with attached delivery boxes. The heatmap reveals concentrations of food delivery workers primarily in Manhattan and parts of Brooklyn during lunch hours, demonstrating how surveillance could target this vulnerable group.}
\end{figure}

We also generate inferences on several other group identities using the zero-shot method from \autoref{tab:adversarial-methodology}, including nurses, religious-presenting individuals, dog owners, police officers, and protesters, all shown in \autoref{fig:teaser}.

\subsubsection{Pentest: Mobile Food Vending and Delivery in New York City}

As of October 2024, Hunter College's Food Policy Center reports that the NYC Department of Health has issued around 4,600 food cart permits and about 500 food truck permits~\cite{nycadmin_new_2014}. Cart permits are in high demand: Thousands of individuals are currently on the NYC Department of Health permit waiting list~\cite{stewart_can_2024}. The current population of food delivery workers in New York City is estimated to be between 65,000 and 122,000, many of whom are immigrants~\cite{noauthor_new_2024, attanasio_getting_2024}. Given their vulnerability in real-world scenarios, we focus our exploration on mobile vendors and delivery workers, a group frequently targeted by adversaries. In April 2024, a group of vendors in Brooklyn protested a surge in ticketing, alleging unfair targeting by the NYPD~\cite{stewart_can_2024}. Just a few months later, in July, police detained a vendor who collapsed while handcuffed and required hospitalization~\cite{stewart_can_2024}. In 2023, the NYPD took street vendors to criminal court almost six times as frequently as in 2019, issuing over 1,200 criminal summonses~\cite{chu_nypd_2024}. Food delivery workers are also targeted by various groups. As early as 2021, The New York Times reported on the theft of workers' e-bikes (valued at up to 3,000 USD), highlighting the need for self-organization~\cite{marcos_as_2021, formoso_bronx_2024, ojeda_newly_2024}. More recently, the NYPD has cracked down on illegal mopeds, with much of the focus directed at delivery workers~\cite{healy_how_2024}.

The 23,000 street vendors in New York City are a particularly vulnerable group with the majority of them vending as their primary source of income. As previously mentioned, these are predominantly immigrants (96\%), and most of them operate in legally gray areas~\cite{settle_street_2024}. Indeed, around 75\% of mobile food vendors have no permit~\cite{settle_street_2024}. While providing a valuable service to New Yorkers by offering fast, convenient, and affordable goods, these small businesses operate in a precarious environment, their continued existence reliant on the challenges faced by enforcement agencies. While exact demographic information on food delivery personnel is opaque, we know that food delivery apps like Uber Eats and DoorDash rarely comply with governmental requests for information~\cite{noauthor_new_2024}. 

Our pentest investigates whether adversaries, such as police, can gain cheap, easy, and more direct access to information about where and when vendors operate using DSI datasets, even when individuals are de-identified. Such information poses a significant threat to the livelihood of these vendors, who already face minimal job security and ongoing concerns about policing~\cite{urban_justice_center_street_2023}. Thus, significant threats to privacy result from the identification of vendors at the group and community level. 

\subsubsection{Pentest Execution}
Having acquired data representing inferred detections of mobile food vending and food delivery workers, we now behave as an adversary might, and attempt to measure the spatiotemporal distribution of these groups in New York City. We labeled the training data needed for the object detection model in 4 hours. We trained a high-performance, convergent model in 44 hours. Using this model, we were able to infer 192 images per second on a single GPU, processing the entire dataset of \numDashcamImages images in just 36 hours. Under an optimal confidence threshold of 0.205, the model asserted 196,183 images depicting food trucks. We show a zoom-in of the Jackson Heights area of Queens in New York City (see Figure 4 in Supplement), a known hotspot for unlicensed vending~\cite{noauthor_vendors_2024}.

\subsubsection{Pentest Findings}
From Experiment 1 we find that out of all studied food truck vending violations\footnote{We assemble a list of all food truck vending violations during our dataset's coverage period from NYC OpenData, specifically the NYC Office of Administrative Trials and Hearings case status dataset~\cite{nycopendata_oath_2025}.}, the median distance to the nearest high-confidence model detection is only 127 feet and gets as near as 36 feet. We find a clear visual overlap between known food truck vending violations and our high confidence detections, shown in \autoref{fig:food-truck-map}. This means that, for areas with high concentrations of food truck vending violations, there is more than ample imagery with which an adversary could pursue remote inspections, similar to the methodology in~\cite{shapira_dorin_fingerprinting_2024}. 

From the detections of food delivery workers in Experiment 2, we are also able to easily create targeted deployment zones for the in-the-wild surveillance of food delivery workers, shown in \autoref{fig:food-delivery-heatmap}. These results demonstrate the ease of developing a useful attack tool against groups with DSI, even under commercially-standard de-identification. 

We have shown the \textit{ease} of crafting useful attack tools. The question remains as to \textit{what} research applications of DSI are ethical to pursue. We tackle this question in the paper's latter sections, using the contextual integrity framework. We now present a background and related work section, and then move to our analysis of information flows in DSI, followed by discussion.

\section{Related Work}
This section describes related and contextual work, including an overview of DSI-producing technologies, the inferential power of computer vision models, group privacy theory, and the framework of contextual integrity.

\subsection{DSI-producing Technologies}
Earlier sensing technologies like Google Street View (GSV)~\cite{anguelov_google_2010} have been used to computationally characterize longer-term societal processes, including gentrification~\cite{thackway_implementing_2023, ilic_deep_2019}, street safety~\cite{naik_streetscore-predicting_2014}, health outcomes~\cite{nguyen_using_2019}, shade from street trees~\cite{thackway_implementing_2023}, and demographic distributions~\cite{gebru_using_2017}. People and people-descriptive objects are visible in these images. Advances in tools that collect street view imagery have attracted considerable scrutiny, notably as facial features~\cite{geissler_private_nodate}, vehicle license plates \cite{elwood_privacy_2011}, and individual homes~\cite{elwood_privacy_2011} become characterizable. As empirical research using GSV has evolved, so has research focusing on pedestrian obfuscation~\cite{kunchala_towards_2023, uittenbogaard_privacy_2019, flores_removing_2010, nodari_digital_2012} and commentary on the ethical use of GSV~\cite{bader_promise_2017, helbich_use_2024}. 

Street view imagery has been used in attempts to characterize short-term processes, such as counting public pedestrians~\cite{campanella_people-mapping_2017, yin_big_2015} and alcohol consumption~\cite{clews_alcohol_2016}. However, the temporal variability of GSV, averaging \textit{7 years} on a controlled, short-route surveying study~\cite{kim_examination_2023}, means it is unsuited for capturing short-term, real-time or near real-time phenomena related to groups. Audit works at FAccT have shown the insuitability of GSV and similar technologies for measuring abstract concepts like 'livability'~\cite{alpherts_perceptive_2024}, particularly when human annotators are involved~\cite{zamfirescu-pereira_trucks_2022}. 

We qualify DSI-producing technologies as those that produce spatially representative data at a frequency sufficient to capture short-term phenomena including people and their behaviors. DSI-producing technologies include networks of static traffic cameras~\cite{dietrich_seeing_2023, snyder_streets_2019} and collections of mobile cameras ("dashcams"), either on private~\cite{franchi_detecting_2023, franchi_towards_2024} or public~\cite{redmill_automated_2023, sandhaus2024crashdata} fleets. The sensing capability of a DSI-producing technology can be evaluated quantitatively as the fraction of all possible space-time pairs with some number of images. DSI can be deployed to fingerprint phenomena at very high granularity~\cite{franchi_towards_2024}. As an example, DSI permits the continuous monitoring of foot traffic patterns on a local sidewalk network at 15-minute increments~\cite{franchi_towards_2024}. 

In the following, we provide background on the key privacy concepts we apply for our analysis of DSI. To account for the complexity of privacy vulnerabilities in DSI, we ground our analysis along three core concepts: inferential models, group privacy, and contextual integrity (CI). 

\subsection{The Inferential Power of Vision Models}
Ascendant inferential capabilities of AI models, particularly computer vision models, pose a significant threat to group privacy in DSI. AI's capacity to draw inferences enables the extraction of specific information of interest without directly collecting it. Computer vision AI `makes sense' of a sea of visual data~\cite{bhatt2021cnn} by drawing inferences from it: vision models in robotics~\cite{cavallo2018emotion}, self-driving vehicles~\cite{janai2020computer}, and emotion recognition systems~\cite{hassouneh2020development} analyze the semantics of images and videos—millions of which are generated daily across different digital socio-technical systems, including the DSI. AI inferences present significant challenges to conceptions of privacy, both in theory and in data protection practice. The privacy conceptions most susceptible to erosion by AI's inferential power are likely those grounded \textit{exclusively} in categorical distinctions -- such as classifying data as sensitive or non-sensitive -- while at the same time uncritically accepting AI-generated inferences as inherently valid~\cite{engelmann2025countering}. Critical data scientists, particularly members of the FAccT community, have demonstrated the adverse impacts of invalid inferential models, especially due to biased misrepresentations that result in improper and unfair predictive descriptions of individuals they cannot understand, correct, or control (e.g.,~\cite{geiger2020garbage, goldenfein2019profiling, ullstein2024attitudes, ullstein2022ai, engelmann2022people, stark2021ethics, engelmann2024visions}).

Although blurring may prevent the processing of facial data from individuals, such approaches do not fully protect against the inferential power of computer vision models in identifying group-relevant attributes. DSI serve both as a training ground for model development and as a deployment environment for trained models. Inferential models can leverage the semantics inscribed in urban centers, buildings, and public squares, which manifest in corresponding scripts of activities, behaviors, and roles. People may gather for protests in public squares, work in specialized buildings such as hospitals or construction sites, or prepare to engage in religious practices outside designated spaces such as temples, churches, synagogues, or mosques. Models can detect such environmental cues and infer sensitive information about facially obfuscated individuals by analyzing group membership proxies such as clothing, accessories, and behaviors. Computer vision models effectively analyze visual data, enabling the inference of clothing types, styles, and the presence of accessories such as bags, hats, glasses, and jewelry~\cite{cheng2021fashion, bossard2013apparel}. This capability facilitates real-time identification of individuals belonging to specific groups, including demonstrators, religious congregations, or professionals (e.g., doctors and nurses), based on distinctive attire and accessories. Even when individuals do not explicitly display group affiliation through their own clothing or accessories, their physical proximity to those who do can result in their association with the group—a phenomenon often referred to as the ``lookalike'' effect~\cite{pyrgelis2017knock}. In public spaces, computer vision models can also infer protected attributes such as gender through pose estimation~\cite{kastaniotis2013gait}, or physical and mental disabilities through proxies such as wheelchair or white cane use~\cite{rosero2018intelligent}. When deployed on the DSI, AI's purported inferential power generates unprecedented privacy challenges, turning public spaces into arenas for real-time, automated meaning-making that facial or body obfuscation alone cannot prevent.

\subsection{Privacy as Appropriate Information Flow vs. Privacy as Preference}
In motivating our choice to utilize the contextual integrity framework in our analysis of information flows in DSI, we now detail two prominent takes on privacy: privacy defined by group preferences (aligning with the theory of group privacy \cite{Taylor2017-sc}), and privacy defined by appropriate information flows (aligning with the framework of contextual integrity \cite{nissenbaum_privacy_2004}).

\subsubsection{Group Privacy}
Group privacy has been the subject of scholarly discussion since the late 1990s, when researchers began examining how new information technologies classify individuals according to shared attributes rather than treating them as isolated subjects. In his prescient work,~\citet{Vedder1999-ui} introduced the concept of ``categorical privacy'' to address how data mining techniques lead to the ``deindividualization of the person,'' where judgments about individuals are based on group characteristics rather than individual merits. This early Dutch work on group privacy emphasized that privacy concerns extend beyond individual data subjects to entire classes of people, anticipating the challenges posed by modern big data analytics.

Building on this foundation,~\citet{Taylor2017-sc} define group privacy as the collective ability of a group to control its personal and shared information. In practice, this involves protecting both individuals' information within the group and safeguarding shared information pertaining to the collective entity itself. This definition recognizes that groups, as distinct entities, have privacy interests beyond those of individual members, considering the shared norms and values that shape collective privacy practices~\cite{floridi_group_2017}.

\citet{bloustein_individual_2017} further developed the concept of group privacy as protecting confidential information shared among two or more individuals against external parties. They outline the ``right to huddle,'' referring to a group's ability to gather and communicate confidentially within their own boundaries, enabling groups to maintain trust, collaboration, and collective decision-making without undue external surveillance or interference. Within our paper, we specifically focus on normative groups observable in DSI, which we discuss in detail in~\autoref{tab:group_types}.

\citet{Loi2020-nr} distinguish between two concepts of group privacy: ``what happens in Vegas stays in Vegas'' privacy, concerning confidential information shared within a group, and ``inferential privacy,'' dealing with inferences about groups defined by shared features. Our pentest results directly implicate this second form of privacy, showing how readily group memberships can be inferred from seemingly anonymized DSI data. \citet{Mantelero2017-bm} further argues that in the context of big data analytics, privacy and data protection should be considered collective rights rather than purely individual ones.

\citet{van-der-Sloot2017-qp} explores whether groups should have a right to protect their group interest in privacy, noting that while privacy rights have historically focused on individuals, contemporary technological paradigms like big data present threats that materialize at group levels rather than individual ones. As~\citet{Asgarinia2024-cs} argues, the traditional focus on individual privacy rights fails to address the vulnerabilities of ``clustered groups'' designed by algorithms, where information about the group can be used for harmful purposes even when individual members remain anonymous. These perspectives highlight the inadequacy of individual-focused privacy frameworks in addressing collective privacy challenges posed by modern data analytics.

As our penetration test demonstrated, traditional approaches to privacy that focus on individual anonymization fail to prevent group-based privacy violations in DSI. While individual privacy frameworks might emphasize control and consent, they prove inadequate in addressing the collective, inferential privacy challenges posed by dense spatial-temporal imagery and AI analysis. This limitation points to the need for a more holistic framework that can address both individual and group privacy concerns in the context of DSI.

\subsubsection{Contextual Integrity (CI)}
The inferential reality of computer vision AI models and real-time DSI produce privacy vulnerabilities for groups in public spaces. For the purposes of our work, Nissenbaum's theory of \emph{Privacy as Contextual Integrity (CI)} helps distinguish legitimate from illegitimate information flows according to contextual norms for such groups~\cite{choksi2024privacy}. Drawing on social theory, social philosophy, and law, CI conceives of social life as comprising distinct social domains or \emph{contexts}, such as commerce, education, finance, healthcare, civic life, family, and friends~\cite{nissenbaum_privacy_2004}. A CI context is ultimately defined by its ends, aims, or goals, which further determine its role in society at large, as well as its values, be it equality, justice, or individual autonomy, among others. As such, in a healthcare context, for example, the goal or aim is to cure and prevent illness, alleviate pain, and there is a commitment to ethical values such as equity and patient autonomy. The precise composition of ends and values may differ across societies, and may even be open to political contestation, e.g.,~in an education context, it is open to debate whether the goals are to enlighten or train, to teach rote skills or encourage creativity, or to generate workers as opposed to enable a responsible citizenry. CI shifts away from notions of privacy as information control or secrecy, and conceives of privacy as the appropriate flow of information: flow that conforms with contextual informational norms. Contextual informational norms define acceptable data practices and may range from implicit and weak---social disapproval of friends betraying confidences---to explicit and embodied---laws protecting journalists refusing to name sources or requiring physicians to maintain the confidentiality of health data. A complete statement of a contextual informational norm provides values for five parameters: data subject, data sender and data recipient (collectively referred to as \emph{actors}), information type (topic or attribute), and transmission principle (the conditions under which information flows)~\cite{nissenbaum_privacy_2004, nissenbaum_contextual_2011, nissenbaum_contextual_2019}.

{
\begin{table*}[t]
\caption{Typology of Identifiable Groups}
\label{tab:group_types}
\centering
\begin{tabular}{p{0.15\linewidth}p{0.5\linewidth}p{0.28\linewidth}}
\hline
\textbf{Group Type} & \textbf{Description} & {\textbf{Examples}} \\
\hline
 \textbf{Self Organized}* & Groups formed voluntarily by members who share a common purpose, set of values, or specific goals. These groups often emerge organically through shared interests or collective aspirations, and function independently of external mandates or authority. & Protesters organizing for a cause, local community action groups, religious (i.e. church or temple) groups. \\

\textbf{Role-based}* & Groups composed of individuals acting within defined roles tied to their social, professional, or communal responsibilities. Membership is typically based on one's job, function, or societal duty rather than personal traits or voluntary affiliation. & Nurses, police, construction workers, tourists, school children. \\

\textbf{Clusters} & Spatially proximate groups of individuals who are temporarily assembled or gathered in a shared physical location, often without preexisting social connections or enduring relationships. Such groups are usually context-dependent and formed by situational proximity rather than shared purpose. & People waiting at a train station, individuals standing in line, commuters at a crosswalk. \\

\textbf{Attribute-based} & Groups organized around intrinsic, often immutable characteristics or shared traits of individuals. Membership in these groups is typically defined by demographics or identity markers, which may influence societal perceptions. & Age groups (e.g., seniors, children), gender-specific communities, racial or ethnic groups, individuals with disabilities. \\

\multicolumn{3}{l}{\footnotesize * Indicates normative groups.}
\end{tabular}
\label{tab:typology}
\end{table*}
}

Actors (subject, sender, recipient) are labeled according to contextual capacities or \emph{roles}, such as physicians, educators, or political figures. Information types are defined according to contextual ontologies, such as an educator's reports about a student's learning progress in an educational context. Transmission principles are the conditions or constraints under which a particular information type flows from senders to recipients. Transmission principles include confidentiality, reciprocity, consent or mandated by law, among others. 
CI (and therefore privacy) is achieved or preserved if all information flows within a particular context align with entrenched informational norms. Hence, to determine the appropriateness of an information flow, one must determine all five parameters characterizing such flow.

Unlike privacy-as-preference approaches, which focus primarily on individual control and consent, contextual integrity offers a more comprehensive framework for analyzing the group privacy harms revealed in our penetration test. CI enables us to evaluate the appropriateness of information flows in DSI by considering not just who is depicted, but how that imagery is collected, processed, and shared within specific contexts. This makes CI particularly suited to addressing the complex privacy challenges posed by DSI, where individual de-identification proves insufficient to protect group privacy. In the following section, we apply the CI framework to demarcate appropriate and inappropriate information flows in DSI based on our empirical findings.

\section{Applying CI: Demarcating Information Flows in DSI}
\sloppy %
\raggedbottom

Motivated by the group identifiability risks posed in commercial DSI data~\cite{corvee_people_2012, leach_detecting_2014}, even under intense de-identification, we provide recommendations rooted in a more holistic approach. Under the contextual integrity framework~\cite{nissenbaum_privacy_2004, barth_privacy_2006}, information flows consist of five components: a subject, a sender, a recipient, an information type(s), and a transmission principle. In applying the CI framework, we dissect information flows within DSI and outline the typical values assigned to each of the framework's five components. 

The \textbf{subjects} in DSI are \emph{groups} of pedestrians. We delineate the typology of identifiable groups in DSI in~\autoref{tab:typology}. In many cases, the \textbf{sender} is the \textit{data provider}. While a vehicle driver is also complicit in the act of the data capture, it is the data provider who makes the decision of when to take an image, how many images to take, and how many images to upload for downstream transfer. Alternatively, an adversarial machine learning model may also assume the role of data provider if it sends its generated outputs to a recipient. Subsequent analysis, such as drawing inferences on top of a given dataset, create a novel information flow, and typically involve a new sender, such as an academic researcher or organization.

\begin{figure*}
    \centering
    \includegraphics[width=\textwidth]{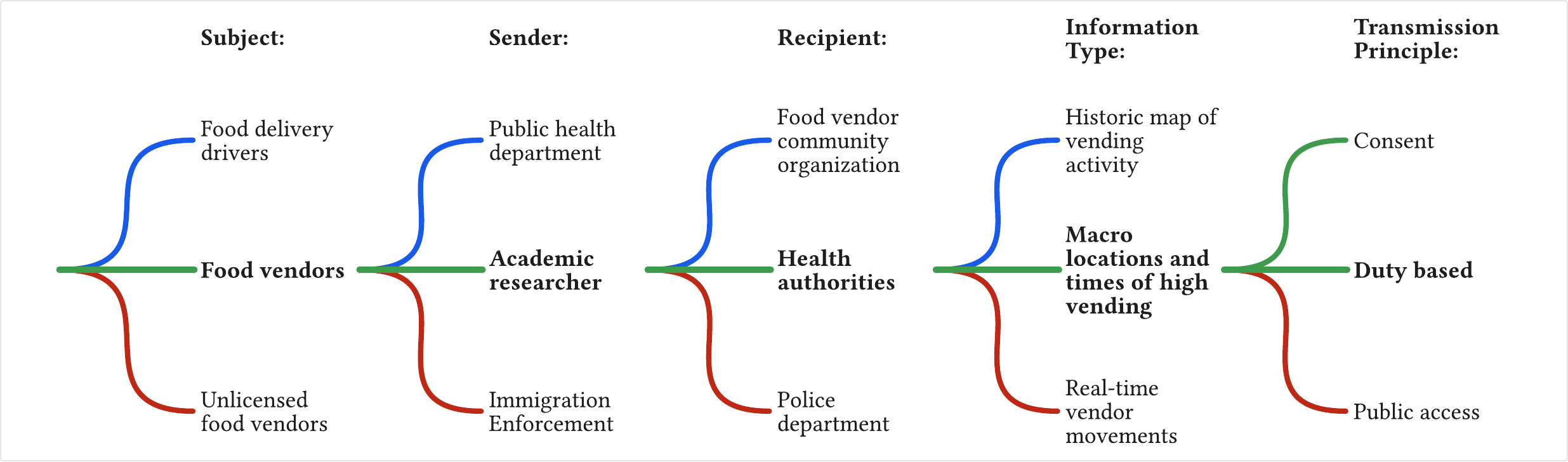}
    \caption{Contextual Integrity Analysis of a DSI Information Flow. Changing a single parameter in an information flow can transform it from appropriate (green, \textcolor[HTML]{3E9B4B}{$\curvearrowright$}) to inappropriate (red, \textcolor[HTML]{BD2915}{$\curvearrowright$}). Contextual integrity requires evaluating fully specified information flows to avoid ambiguous cases (blue, \textcolor[HTML]{185AE9}{$\curvearrowright$}).}
    \Description{Contextual Integrity Analysis of DSI Information Flows. This figure illustrates how changing a single parameter in an information flow can transform it from appropriate (green) to inappropriate (red). The central appropriate flow (all green connections) shows food vendors' macro-level data flowing from academic researchers to health authorities under duty-based principles. Each variation demonstrates how contextual integrity requires evaluating complete information flows rather than individual parameters in isolation.}
    
    \label{fig:DSI-flow}
\end{figure*}

The \textbf{recipient} is variable. In most research projects that use DSI to date, the recipient is a \emph{research group}. DSI can also have commercial uses, in which case the recipient is a \emph{private company}, or a \emph{public sector agency}. In DSI, \textbf{information types} are \emph{photographs} with attached geospatial telemetry data. This combination creates a record of a group's (or group member's) location and the time they were situated there. DSI involves a \textbf{transmission principle} where the subjects are not required to give consent and have no right to revoke the transmission unless they preempt the information flow via requesting an obfuscation of their appearance in the dataset.\footnote{Google supports identifiable content blurring in Street View~\cite{noauthor_blur_nodate}, but not removal. The European Union’s GDPR framework advances the right to removal through its articulation of the ``right to be forgotten.''~\cite{politou_forgetting_2018}.} As we demonstrate in \autoref{fig:DSI-flow}, the contextual integrity framework provides a structured approach to evaluate whether information flows in DSI respect privacy norms by examining the complete five-parameter tuple rather than isolated elements.

\subsection{An Inappropriate Data Flow in DSI}
According to CI~\cite{nissenbaum_contextual_2011}, data flows that breach socially accepted norms are considered 'inappropriate.' DSI, in combination with automated image retrieval through algorithmic methods (described in \autoref{tab:adversarial-methodology}), radically disrupts many contextual information norms, similar to the advent of big data technologies~\cite{Lane2014-ns}.    

As an extension of our penetration test, we consider food delivery workers as a data subject. Prior to the introduction of street view and DSI technologies, encounters of food delivery workers and authorities in urban environments were ephemeral. Records of food delivery were noticed by the local community or public, in journalism, media, and writing, or within individually distributed photographs. In these settings, food vendors could reasonably react to the presence of authority, and communicate in person.

In this context, we conceptualize the city as the recipient (or adversary) of the data, with the data provider acting as the sender. The data itself consists of images enriched with computational metadata, including precise latitude and longitude coordinates, as well as the exact date and time of capture. Crucially, this transmission is sufficient to track a vendor’s activities over time or issue fines for operating without a license. The underlying transmission principle assumes that vendors cannot revoke the transfer of this data, nor are data providers required to inform them that they have been recorded.

A proponent of DSI surveillance might argue that DSI providers must give the police persistent access to food vendors' vending locations and movement throughout the city upon request. While this information flow may, at face value, appear morally justifiable, it causes direct privacy harms to food vendors. Such surveillance exposes vendors to risks of increased fines, job loss, or even the inability to continue their work, threatening their livelihood and way of life. This impact is particularly severe for undocumented immigrants, for whom food vending is not only a direct lifeline to sustain their families~\cite{marcos_as_2021}.
Disrupting this flow has moral implications such as economic survival, access to low-cost food for others, and respecting the right to work without unnecessary or harmful interference.
 
\subsection{Inappropriate Flows in Other Groups}
We highlight examples of inappropriate information flows that arise as DSI technologies become more pervasive. These examples, some of which are illustrated in our dataset (\autoref{fig:teaser}), offer a glimpse into the many ways DSI and inferential models can violate established privacy norms, posing risks to the interests and values of individuals and groups in public spaces.
Using CI, we trace the following components of each information flow: (1) the subject, (2) the sender, (3) the recipient, (4) the information type(s), and (5) the transmission principle. We specifically locate groups in the DSI imagery that pose contextual integrity \emph{harms} from merely being present in a geo-temporally tagged image. That is, there are documented, real-world instances of a group being targeted based on their public situation in space and time. Below we describe several complete flows for groups from the typology, involving plausible harms to those groups (this information is also presented in a table in the appendix~\autoref{tab:group_types}). 

\emph{Protesters and demonstrators:} Protesters' meeting locations and times, captured in geo-tagged images by DSI providers, may be surreptitiously shared with political groups opposed to their cause. This information flow enables adversarial groups to target and disrupt peaceful assemblies, undermining the protesters' right to organize and express dissent. The resulting harm includes exposure to retaliation, suppression of free speech, and the erosion of democratic principles. \emph{Workers and employees:} In professional settings, shift patterns of nurses inferred from DSI images captured outside healthcare facilities could be sold to exploitative employers or staffing agencies seeking to take advantage of their availability. Such data flows enable adversaries to target nurses with aggressive recruitment tactics, pressure them into accepting poorly compensated shifts at inconvenient hours, or manipulate them into working under unsafe or undesirable conditions. This undermines nurses' autonomy in the workplace and their ability to make independent decisions about their labor. \emph{Pedestrians:} Geo-tagged imagery of people gathered at busy crosswalks during peak hours may be sold by predatory advertisers seeking to exploit behavioral patterns. For example, advertisers might use this data to push high-pressure marketing campaigns for products like payday loans or fast food, directly targeting the commuters' mobile devices in certain locations or interactive billboards. \emph{Religious groups:} Muslims or Hasidic Jews, photographed in locations perceived as incongruent with their practices (e.g., near entertainment districts), risk having these images and their associated metadata shared with hate groups, potentially exposing them to targeted harassment or discrimination. This flow facilitates targeted harassment, stigmatization, and even violence against the group, violating societal norms of religious freedom.

\subsection{Appropriate Data Flows in DSI}
While many data flows enabled by DSI can lead to privacy violations, this technology also has beneficial uses, and privacy can simultaneously be protected when contextual norms are respected. We briefly outline two examples of appropriate information flows that align with societal expectations and provide public benefit.

\emph{Urban planning for pedestrians:} City planners can use DSI to study pedestrian movement patterns at crosswalks and intersections to optimize traffic signal timing and improve infrastructure. Transit authorities can analyze DSI, showing commuter congregation patterns at bus stops and train stations to adjust service frequency and capacity.

\emph{Crisis support for homeless:} Non-governmental organizations receive DSI to study the temporal movement of encampments. This allows crisis teams to provide on-demand support to those communities in need. 

\autoref{fig:DSI-flow} illustrates how contextual integrity operates in the scenario described throughout this analysis. The center pathway represents an appropriate flow: academic researchers sharing macro-level vending patterns with health authorities under duty-based principles. However, altering a single parameter, such as changing the recipient to law enforcement or modifying the information type to real-time movements, transforms the flow into an inappropriate one that violates privacy norms or an ambiguous one. The transmission principle outlines conditions under which data is obtained, used and reused, for example, the same vending location data that enables public health planning becomes problematic when made publicly accessible without restrictions.

These examples demonstrate how DSI can be deployed ethically when the information flow of DSI and its inferences are fully specified according to CI, as well as when: (1) The purpose serves a clear public benefit, (2) Usage is limited to the legitimate use purpose, (3) Access is restricted to appropriate and previously agreed parties, (4) Affected groups are given agency in how their information is collected and used, (5) Focus is restricted to those groups by limiting inferrable information.

The distinction between appropriate and inappropriate information flows hinges not only on the type of data collected or who the data subject is, but on the full specification of the flow across all contextual integrity parameters. To ensure DSI technologies are used responsibly, robust safeguards and ethical guidelines must be implemented to protect privacy while enabling beneficial applications.

\section{Discussion \& Concluding Remarks}
We conclude by returning to the targeted surveillance of mobile food vendors in New York City. Combined with findings from our penetration test, we demonstrated that DSI can effectively be used to identify contextual information about group distributions and facilitate harm against group members. The object detection model generated (photograph, place, time) tuples for tens of thousands of food trucks from any given set of images. At present, the most realistic threat model for these outputs is largely \textit{curation}. Authorities, however, with additional information on food truck congregations and outlying locations, can ramp up targeting in those areas, likely resulting in increased fines and summonses \cite{stewart_can_2024}.

\subsection{Issue of Reducing Privacy to Anonymity}
This work challenges the assumption that anonymizing individuals within DSI sufficiently protects privacy. Through contextual integrity analysis, we identify group vulnerabilities created by inferential model development, illustrated in \autoref{tab:group-harms}. Despite being promoted as effective protection against privacy harms \cite{chaum_untraceable_1981, matthews_anonymity_2010}, anonymity fails against AI's inferential capabilities. Our penetration test shows facial blurring provides no protection for food delivery workers against adversaries analyzing group membership. Authorities accessing DSI could organize targeted enforcement leading to loss of income, imprisonment, and other negative consequences.

The increasing density of street imagery, combined with advances in image retrieval techniques, makes it possible to circumvent anonymity in practice. As~\citet{nissenbaum_big_2014} explains: ``Even where strong guarantees of anonymity can be achieved, common applications of big data undermine the values that anonymity traditionally protected. Even when individuals are not `identifiable,' they may still be `reachable' and subject to consequential inferences and predictions made on that basis.'' Anonymity alone is insufficient to safeguard against the broader harms enabled by DSI and inferential AI models. CI's theory of privacy does not deem all information flows involving 'sensitive' data as inappropriate. Instead, it emphasizes the need to evaluate fully defined information flows, which include the subject, sender, receiver, information type, and transmission principle~\cite{shvartzshnaider_analyzing_2018}. This nuanced approach ensures that privacy judgments are context-specific and grounded in the norms governing the particular scenario. For example, a research or citizen advocacy group with access to the same New York City food delivery worker distribution, specifically tied to the use of that data for a legitimate purpose—such as surveying the state of food delivery in the city—would likely not be considered an inappropriate information flow. The primary use of DSI in this example is positioned as delivering social benefits, such as enhancing city planning and improving citizen safety through the understanding of group behaviors. Given these intended benefits, CI is the chosen privacy framework to evaluate and guide the responsible use of such DSI datasets. Full removal of all traces of human activity in DSI, such as by blocking all humans, their attachments, and their vehicles, would undermine the utility of these tools. 

\subsection{DSI's Threat to Public Space Itself}
Beyond harms to specific groups identified in our penetration test, contextual integrity theory highlights a fundamental concern: DSI threatens the nature and value of public space itself as a social resource~\cite{van-der-Sloot2017-qp, Macnish2012-rc}. Public spaces have historically functioned as domains where contextual norms allow for spontaneity, economic opportunity, and democratic participation—values now threatened by surveillance infrastructures.

For food vendors and delivery workers in our study, this transformation is particularly consequential. What historically served as accessible venues for entrepreneurship—street corners and public thoroughfares—become sites of enforcement vulnerability when continuously documented through DSI. As our penetration test demonstrates, the ability to track food vendors' locations and movement patterns fundamentally alters the social contract that has governed public space use. When vendors' presence in a particular location becomes algorithmically flagged (as shown in \autoref{fig:food-truck-map}), "perfect enforcement" becomes a possibility. Unlike before, where food vendors in New York City were able to establish themselves gradually, eventually gaining sufficient community support for advocacy groups to campaign effectively for policy reforms~\cite{settle_street_2024}.

Similarly, for protesters, religious communities, and other groups identified in our typology (\autoref{tab:typology}), DSI fundamentally reconstructs public space from a domain of relative freedom to one of persistent visibility. As Ben Green asserts in \textit{The Smart Enough City}, the unfettered use of technologies like DSI are moving society towards a state where to avoid being tracked, you must take the quixotic step of opting out of public space~\cite{green_smart_2019}. The chilling effect on public assembly, worship, or everyday activities represents not merely a privacy harm to these groups, but a diminishment of public space's societal value.
This perspective suggests that CI can help us understand how DSI may alter the implicit social agreements governing public spaces. Our findings indicate that the increasing presence of DSI, combined with AI analysis capabilities, could shift what~\citet{Lane2014-ns} describe as "reasonable expectations" of contextual privacy in public spaces. As our food vendor case study suggests, these changes may disproportionately affect vulnerable groups who rely on public spaces for essential activities.

\subsection{Recommendations} 
Motivated by the penetration test and the theoretical group privacy framework under contextual integrity, we suggest establishing responsible data practices particularly directed at research institutions and DSI providers. Additionally, we advocate for more nuance in technical approaches to privacy protection in DSI. 

\paragraph{Require DSI dataset usage approvals}We call academic institutions to establish oversight bodies, similar to the Research Ethics Board proposed in the Menlo Report~\cite{bailey_menlo_2012}, to evaluate the ethical implications of research involving DSI datasets, particularly when such work falls outside the traditional scope of Institutional Review Boards (IRBs). Twelve years after the Menlo Report, we notice that our own institution lacks provisions for ethical review outside of research that deals directly with \textit{human subjects}. In fact, prior work that utilizes DSI has been deemed IRB exempt, even when studying an innately societal phenomenon like police deployments~\cite{franchi_detecting_2023}. As DSI and other sensitive datasets that \textit{depict} individuals without their notice or consent are increasingly shared with researchers, we recommend that universities apply greater scrutiny to the projects that use them.

\paragraph{Establishing DSI data usage norms and promises}As DSI imagery becomes more widely available and the cost of associated analytics continues to decline (see our list of potential algorithmic group identification methods in~\autoref{tab:adversarial-methodology}), companies that provide access to these datasets should take responsibility for ensuring their ethical use by researchers, governments, and corporations. Unconditional sharing, without legal repercussions, will inevitably cause privacy harm to groups. We propose that DSI providers adopt a more practical approach to ensure that data sharing is restricted solely to legitimate purposes. Sharing and reuse should only occur under a new \emph{transmission principle}: purpose-limited, privacy risk-assessed, and with usage and reuse documented. To help achieve this, \textit{researchers} working with DSI need to develop frameworks, databases, and a centralized system to track use agreements and ensure accountability in the sharing process. This system could be established as a ``Data Use Agreement Database'' where each use case is logged with respect to its purpose, risk assessment, and compliance with privacy protections.

\paragraph{Study Societal Norms} A crucial aspect of understanding proper information flows in DSI involves examining societal expectations regarding the depiction and inferrability of groups in such datasets, in line with work studying the public perception of DSI-producing technologies like CCTV \cite{mazerolle_social_2002}, dashcams \cite{ashrafi_technology_2024, gruchmann_big_2025}, and smart glasses \cite{kaviani_exploring_2024}. To acquire this knowledge, researchers should conduct factorial vignette studies~\cite{Aguinis2014-tk} to build evidence that helps demarcate appropriate and inappropriate flows~\cite{Bourgeus2024-ub} in this novel technology. We propose research that puts a strong focus on potential privacy harms to groups.

\paragraph{Develop contextual obfuscation tools for DSI}There is a pressing need for more nuanced privacy protection tools tailored to DSI. As our pentest demonstrates, good faith privacy protections like facial de-identification are insufficient in protecting group privacy and, in certain instances, individual privacy too. A more robust approach such as blurring entire bodies and vehicles within a DSI frame may offer a short-term solution, but it falls short in two important respects. First, they substantially reduce the image's visual quality and utility. Second, they do not go far enough, as other surrounding factors, like body attachments and color information from blurred rectangles, allow for contextual identification (see \autoref{tab:obfuscation-taxonomy}). In \autoref{fig:nexar-obfuscation-present}, we show how full-body blurring reduces DSI frames' visual quality and utility, while still leaking indicators permitting the inference of a farmer's market event happening. Further, other objects not attached to people can pose privacy threats to groups. This is shown in the supplement in \autoref{fig:food-truck-map}, where we document how inferences of food trucks, despite the blurring of license plates, pose similar privacy risks to food vendors.

In summary, we argue that image blurring techniques must move beyond generic object suppression and begin to treat privacy as a matter of contextually appropriate information flow. This requires the development of contextually aware obfuscation tools~\cite{burdon_object_2024}. Rather than indiscriminately blurring objects, such tools should assess what information about individuals, or groups, might be inferred from contextual cues, and adapt the level of de-identification accordingly.

\subsubsection{Future Work}
In our work, we assume that an adversary is human. However, in contexts where fully automated enforcement targeting a group is implemented, it is imperative to evaluate the performance of leading Vision-Language Models (VLMs) in tasks such as de-identifying and clustering groups. This profiling could yield critical insights into the capabilities and limitations of automated systems in enforcing group-based surveillance or interventions. Further, there is a need for downstream
downstream efforts in academia that promote proactive ethical decision-making in research, especially when working with DSI and other data streams that enable group-level measurement and identification, such as cell phone mobility data.

\clearpage
\section*{Endmatter Statements}
\subsection*{Author Contributions}
All authors discussed the results and implications and commented on the manuscript, but contributed at different stages.\newline
\textbf{Matt Franchi:} Ideation, Conceptualization, Methodology, Software, Formal analysis, Investigation, Data curation, Writing - Original draft, Review \& Editing, Visualization, Discussion. \\
\textbf{Hauke Sandhaus:} Conceptualization, Methodology, Theory, Investigation, Writing - Original draft, Review \& Editing, Data curation, Discussion. \\
\textbf{Madiha Zahrah Choksi:} Theory, Writing - Original draft, Group typology.\\
\textbf{Severin Engelmann:} Writing - Original Draft, Inferences, Review \& Editing.\\
\textbf{Wendy Ju:} Supervision, Resources.\\
\textbf{Helen Nissenbaum:} Supervision, Theoretical oversight.

\subsection*{Acknowledgments}
We thank Hal Triedman, Ricky Takkar, and Tom Ristenpart for formative feedback on a research proposal for this project. We are grateful for generative discussions with our colleagues Ilan Mandel, Gabriel Agostini, Sidhika Balachandar, and Emma Pierson at various stages of our research. We also thank the Digital Life Initiative Reading Group at Cornell Tech, and the Urban Tech Hub at Cornell Tech for longstanding research support. Finally, we thank Nexar, Inc., for collaboration, support, and data access.

\subsection*{Positionality}
Our interdisciplinary research team brings together diverse perspectives and expertise that shaped this work. The author team includes computer scientists with backgrounds in computer vision, machine learning, design, and human-computer interaction; interdisciplinary scholars with expertise in privacy law and data ethics; and philosophers focused on privacy theory and normative technology ethics. 

All of our authors are based at a technology research institution in New York City, giving us firsthand experience with the urban environment we study. This positioning has informed our understanding of how dense street imagery technologies operate in practice and their implications for urban residents. We acknowledge that our institutional affiliations may influence access to resources, datasets, and industry partnerships that facilitated this research.

\subsection*{Ethical Considerations}
In our penetration testing experiment, where we assume the role of adversaries, we are committed to ensuring that no harm arises to any individuals depicted or inferable in our data. As a preliminary measure, all data used in this study was collected during 2023 and is over a year old at the time of publication. Our study exclusively focuses on groups that could face civil repercussions as a result of adversarial detection, deliberately avoiding cases where detection might lead to criminal consequences. To prevent any potential misuse of the models developed during this research, we delete all model weights and checkpoints prior to the project's conclusion. We acknowledge our positionality as researchers and the interdisciplinary nature of our team. Our collaboration is emblematic of a balanced approach to both technical rigor and ethical considerations. Our methodology reflects a commitment to responsible data handling, focusing on the broader implications of DSI without exposing individuals or their specific contexts.

\subsection*{Adverse Impact}
We acknowledge and reflect on the fact that this research may cause readers to experience heightened anxiety over the state of increasing surveillance in our society, especially as we shed light on DSI, a relatively unknown technological product.

\clearpage
\bibliographystyle{ACM-Reference-Format}
\bibliography{manual-sources,group-privacy}

\clearpage
\appendix
\setcounter{figure}{0}
\renewcommand{\figurename}{Figure}
\renewcommand{\thefigure}{S\arabic{figure}}
\setcounter{table}{0}
\renewcommand{\tablename}{Table}
\renewcommand{\thetable}{S\arabic{table}}

\section{Supplemental Figures}
\label{supplement}

\begin{figure}[h!]
    \includegraphics[width=0.48\textwidth]{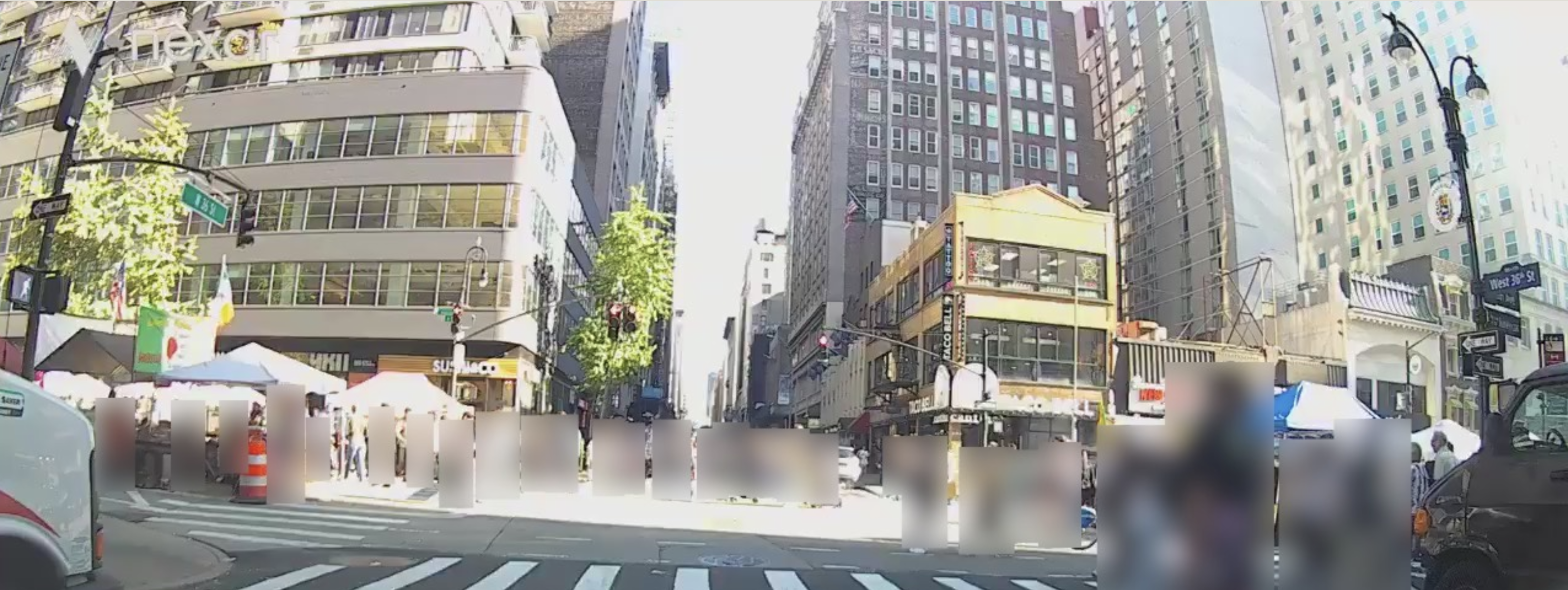}
    \caption{An example of a DSI image processed by Nexar's current pedestrian obfuscation algorithm, showing that pedestrians' entire bodies and faces are obscured by a blurred rectangular box. While individual identities are rendered nearly impossible to infer, environmental cues still allow for the inference of the scene being a farmer's market.}
    \label{fig:nexar-obfuscation-present}
    \Description{A dashcam image showing a street scene with a pedestrian whose face has been blurred using Nexar's standard obfuscation process. The blurring effectively prevents facial recognition while still allowing the general scene context to remain visible.}
\end{figure}

\begin{figure}[h!]
    \includegraphics[width=0.48\textwidth]{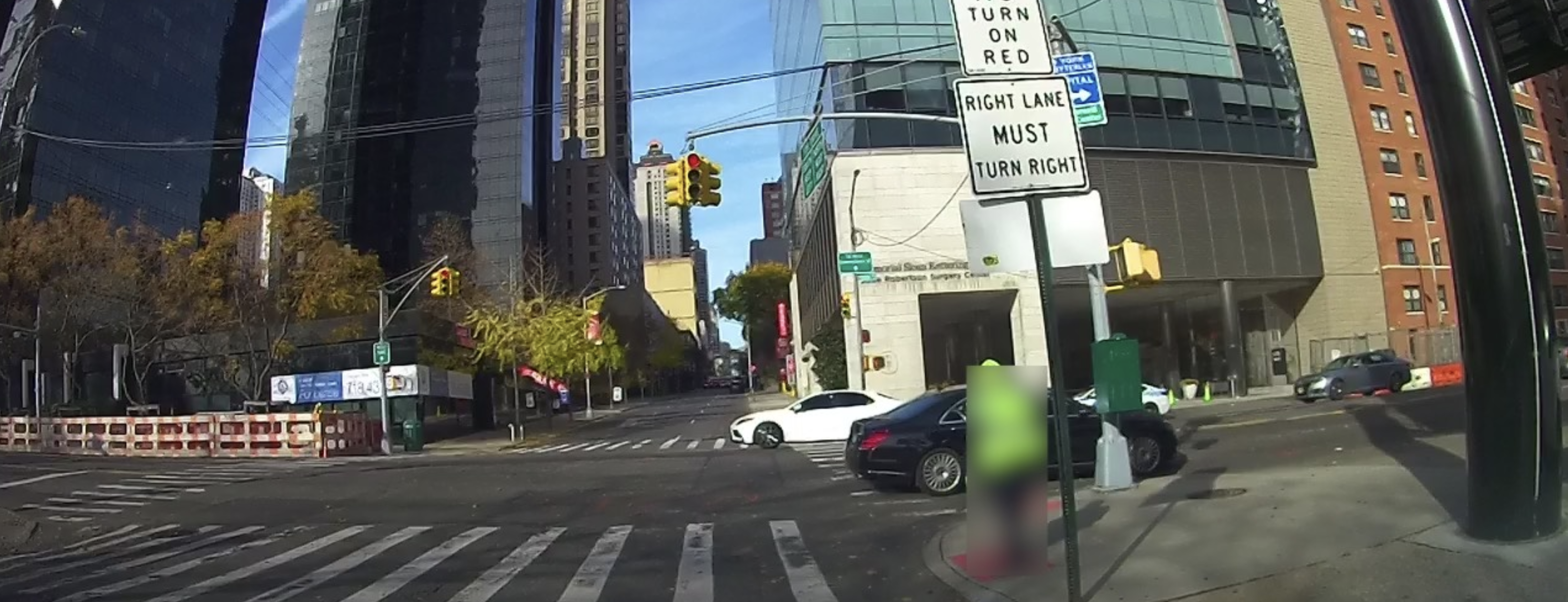}
    \caption{An example of group membership inference, even under full-body pedestrian obfuscation. Due to the high-visibility vest worn by this NYPD traffic officer, a group membership inference can be made solely from neon-green color, black pants, and situation on the corner of a traffic intersection.}
    \label{fig:nexar-obfuscation-highviz}
    \Description{A dashcam image showing a traffic officer wearing a neon-green high-visibility vest, demonstrating how group membership can be inferred despite obfuscation.}
\end{figure}

\begin{figure}[h!]
    \includegraphics[width=0.48\textwidth]{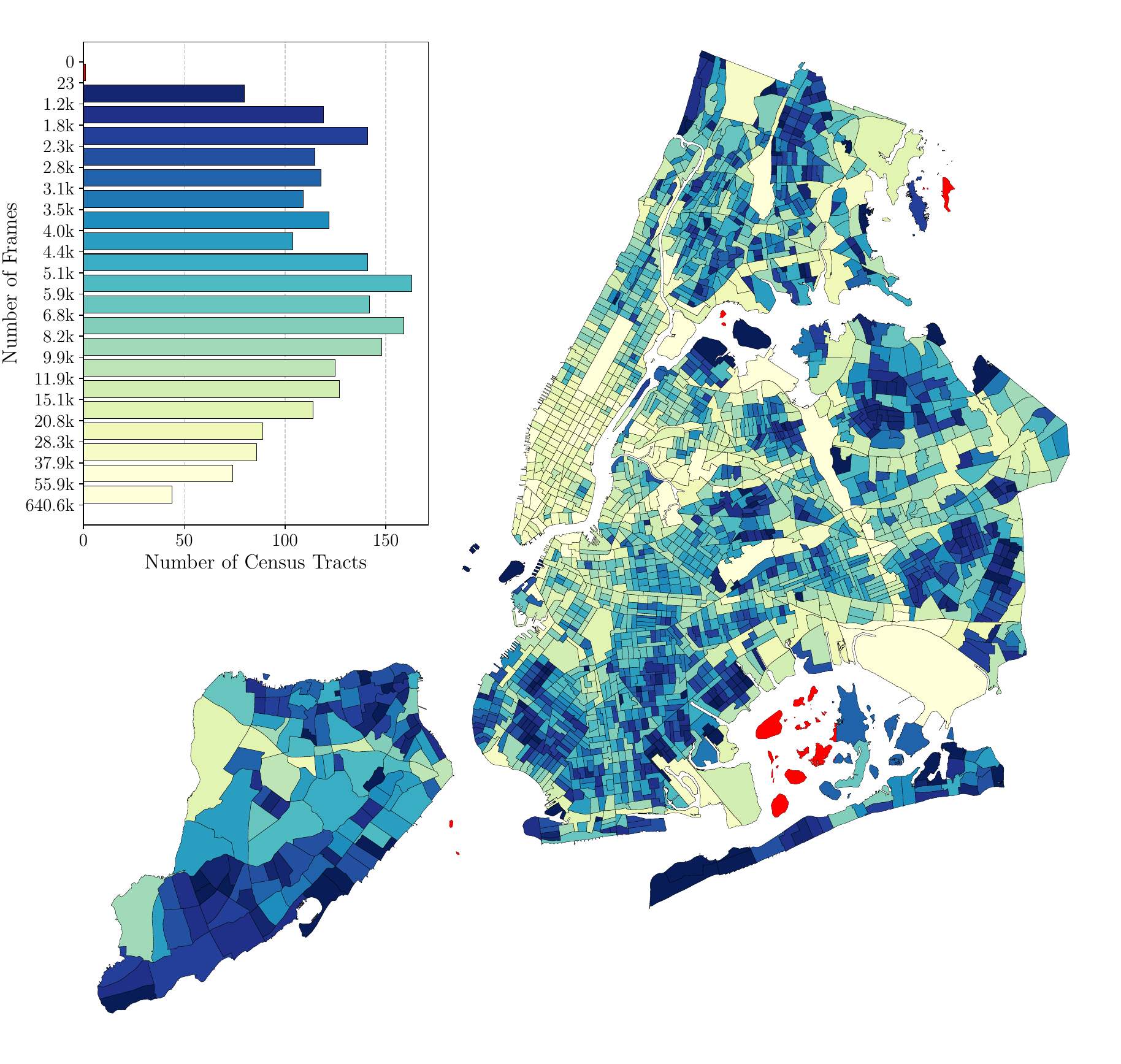}
    \caption{Chloropleth map of the census tracts of NYC, colored by number of dashcam images acquired in a tract. Counts can be referenced in the accompanying frequency histogram.}
    \label{fig:coverage}
    \Description{A chloropleth map of NYC census tracts showing the number of dashcam images acquired in each tract, with a histogram displaying the frequency distribution of image counts.}
\end{figure}

\clearpage

\section{Further Details on Attack Tool Implementations}
\label{sec:B_nitty-gritty}
\subsection{Experimental Design}
We design two experiments to demonstrate the privacy threats that commercially standard DSI imagery can pose to groups when combined with image retrieval and processing using machine learning (ML).

\subsubsection{Sourcing DSI}
We developed software to systematically extract dashcam images, organizing them by date and capture location. Between August 11, 2023, and January 10, 2024, we collect a total of \numDashcamImages images, ensuring comprehensive geographic and temporal coverage.\footnote{We provide additional context regarding our sampling process, which was notably complex. On October 1, 2024, the data provider overhauled the API used for downloading frames and metadata, rendering our tool inoperable. We adapted to these changes and resumed data collection on October 20, 2024. Alongside the API overhaul, the data provider announced a reduction in the daily volume of uploaded imagery, attributing the change to operating cost constraints.} Throughout the sampling period, the data provider maintained an assertion that sampled imagery was driven by (1) crafting a representative sample and (2) replacing stale imagery with fresh imagery. An important and key limitation of this dataset is that we can not independently verify that images were randomly sampled from the network of cameras, nor can we access statistics about the members of the camera network. We only have access to downstream imagery. 

To illustrate the temporal density of the dashcam dataset, we group the images into 15-minute intervals that encompass all of the times that the scraper was fully operational. This produces 6,144 intervals. Out of these intervals, only 4, produce no new imagery; when investigating, we find that this corresponds to the clocks being set back an hour at the end of daylight savings time, on November 5 2023. The mean 15-minute interval produces 3,444 new images, across an area 350.3 square miles (for reference, New York City and its water areas encompass about 469 square miles).\footnote{Here, we calculate the area spanned by all images in a 15-minute interval by computing the convex hull of the image subset.}

We offer an important caveat regarding the rigor of our data validation and training process. Our objective is investigatory rather than focused on developing a highly accurate model. Several reasons inform this approach: primarily, achieving high accuracy would likely necessitate the use of crowdsourced human annotators \cite{franchi_detecting_2023}, whom we are unwilling to expose to sensitive imagery. Second, our core goal is to demonstrate that group privacy threats are both real and present in DSI data. For this purpose, somewhat imprecise distributions are sufficient to support our findings.

\subsubsection{Crafting Data.}
\label{sec:yolo-model-training}
We query approximately 500,000 randomly-sampled images with Cambrian-13B, a leading open-source vision language model \cite{tong_cambrian-1_2024}, in 2.3 days, asking the model for each image, \textit{'Does this image show a food truck?'}. Of our subset, 2,903 are inferred positive, and 557,602 are inferred negative. We then randomly sample 2,000 images from the set of predicted positive images for human annotation. Two authors distributed the labeling process among themselves. We randomly sample 50,000 images from the set of predicted negative images to use as background images in our model. Then, from this set of 52,000 images, we craft a 60-20-20 training-validation-test split; we will use these splits in training a more lightweight object detection model downstream in the penetration test.

\subsubsection{Data Validation}
Cambrian-13B is a time and compute-intensive model; we require a RTX A6000 GPU with 48GB of ram to load and infer images with the model, and inference takes around 2 seconds per image. That said, it has empirically useful zero-shot accuracy. We evaluate Cambrian's ability to classify images as containing a food truck. To evaluate the precision of Cambrian on this task, we randomly sample 2000 images from the set of images classified positive, and manually annotate them. Of these 2000 images, 1496 contain food trucks (true positives), and 645 do not (false positives), yielding a true positive rate (TPR) of 0.70. For false positives, we manually annotate 'decoys' in the image, or objects that we infer Cambrian mistook for a food truck~\footnote{False positives detected by Cambrian stem from two sources. Visual Confusions: Objects that look like food trucks, such as NYC dining sheds with LED signs. Language Confusion: Trucks featuring food images that lead to the incorrect assumption they are food trucks.}. To evaluate the recall of Cambrian on this task, we randomly sample 200 images from the set of images classified negative, and manually annotate them. Of these 200 images, 3 depict food trucks, yielding a false negative rate (FNR) of 0.015 \footnote{It is worth noting that two of the three of these images are partially or almost fully blocked by an full-body obfuscated pedestrian}. This seems small, but the FNR will magnify at the scale of our dataset; we estimate that Cambrian-13B missed 377,136 images with food trucks. Nonetheless, Cambrian is effective at demarcating positives and negatives; effectively, an image that Cambrian predicts as having a food truck will depict a food truck 70\% of the time, and an image that Cambrian predicts as not having a food truck will have a food truck only 1.5\% of the time.

\begin{figure}[h]
    \centering
    \begin{subfigure}[t]{0.3\textwidth}
        \includegraphics[width=\textwidth]{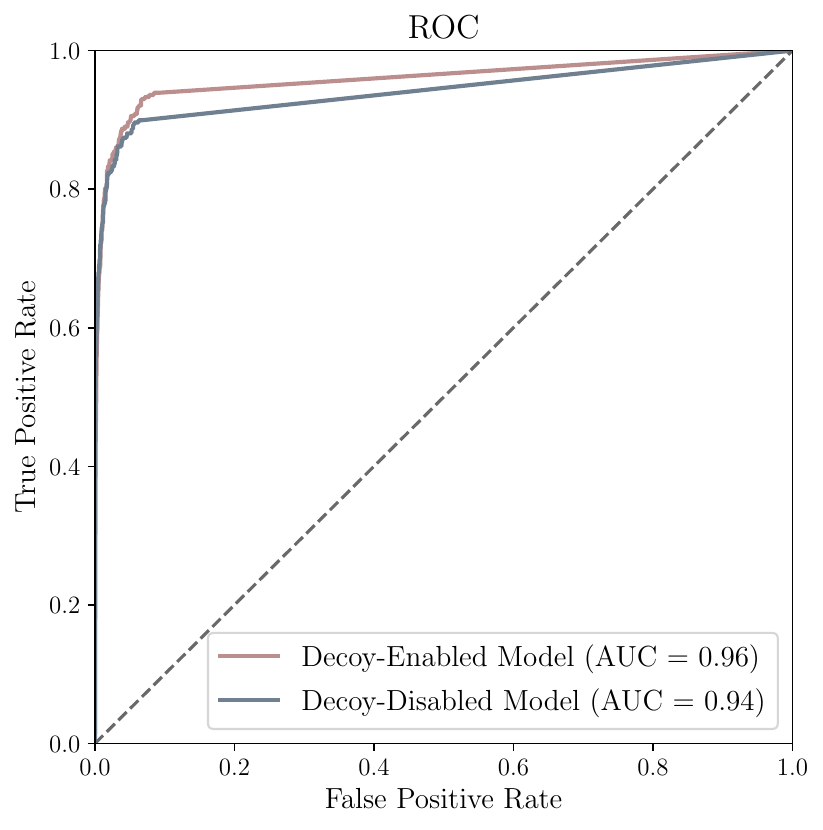}
        \label{fig:ml-roc}
        \caption{Receiver operating characteristic (ROC) curve.}
    \end{subfigure}
    \begin{subfigure}[t]{0.3\textwidth}
        \includegraphics[width=\textwidth]{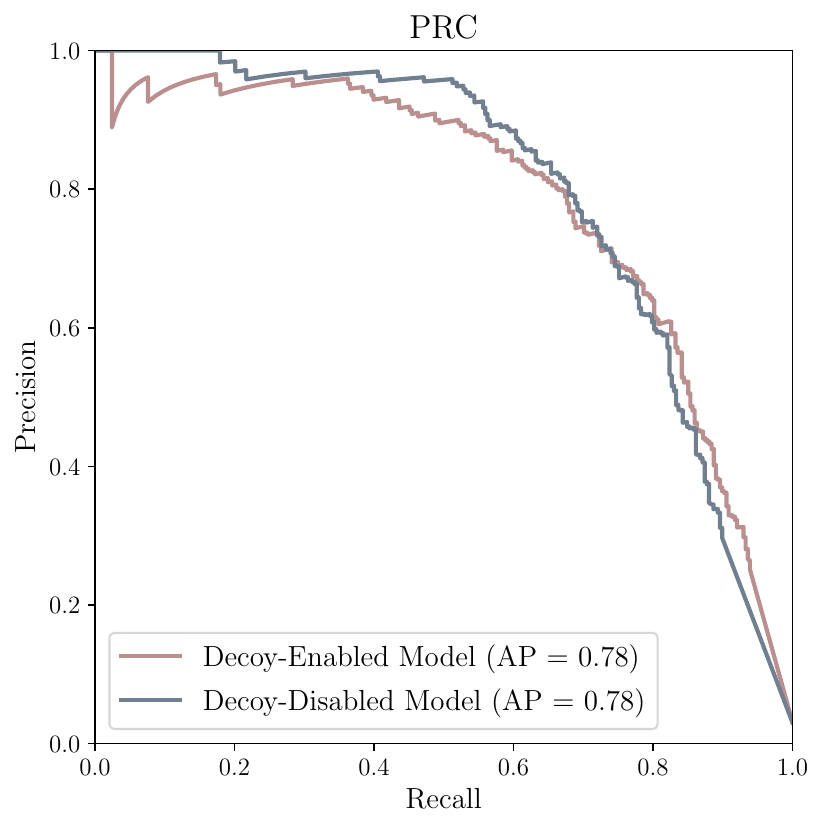}
        \label{fig:ml-prc}
        \caption{Precision-recall (PR) curve.}
    \end{subfigure}
    \caption{Standard performance curves for our food truck object detection model.}
    \label{fig:ml-performance}
\Description{Two performance plots for food truck detection models. The left plot shows the Receiver Operating Characteristic (ROC) curve comparing true positive rates against false positive rates, with curves for models with and without decoy objects. The right plot shows Precision-Recall curves for both models. Both models achieve similar average precision of 0.78, with the decoy-enabled model achieving slightly better AUC (0.96 vs 0.94).}
\end{figure}

\subsubsection{Model Validation}
\label{sec:cambrian-model-validation}
The trained model fits well to the manually-labeled positive images from Cambrian. \autoref{fig:ml-performance} shows the Receiver Operating Characteristic (ROC) and Precision-Recall (PR) curves for the most performant decoy-enabled model and most performance decoy-disabled model, as evaluated on the test set. Both models achieve an average precision (AP) of 0.78. As the decoy-enabled model achieves higher AUC (0.96 vs. 0.94),  we select it for inference on the entire dataset, or "deployment". 

\subsubsection{Lightweight Object Detection Models}
We train You-Only-Look-Once (YOLO, specifically YOLOv11 \cite{jocher_ultralytics_2023}) object detection models on our crafted dataset to induce the capability to identify group members in the entire set of dashcam images. We use the this architecture as it is a standard and principled tool used for object detection tasks in urban scenes (\cite{franchi_detecting_2023}, \cite{franchi_towards_2024}, \cite{shapira_dorin_fingerprinting_2024}). We train one model per object of interest, and report standard performance metrics from the data splits described in the previous section. 
\FloatBarrier

\begin{table*}[t]
    \def\arraystretch{1.15}
\begin{tabular}{lp{10cm}}
\toprule 
\textbf{Mode of Failure} & \textbf{Description of object de-identification issue} \\ 
False Negatives & System fails to accurately detect and blur features of an object before images are made publicly available. \\ 
False Positives & System detects and blurs an object which is not required to be blurred.\\
The "Streisand" Effect & System blurs/blocks an object, and the act of blurring paradoxically draws attention to the object which is intended to be concealed. \\ 
Contextual Identification & System accurately detects and blurs/redacts features, but the object is identifiable from contextual indicators. \\ &\\ \hline
Membership inference & System blurs, blocks or de-identifies several objects of the same class in a visually-similar fashion, allowing for clusters to be generated that leak privacy and enable group membership inference. \\
\bottomrule 
\end{tabular}
\def\arraystretch{1}

    \caption{De-identification failure modes in dense street imagery. The top segment of the table is inherited from~\citet{burdon_object_2024}.}
    \label{tab:obfuscation-taxonomy}

\end{table*}

\section{Taxonomy of De-identification in DSI}
\label{sec:C-obfuscation-taxonomy}
~\citet{burdon_object_2024} identify four key failures in the standard de-identification approach of blurring faces, license plates, and other identifiable objects: false negatives, false positives, the 'Streisand effect,' and contextual identification. False negatives and false positives are common concepts in the machine learning literature, referring to instances where de-identification fails to blur identifiable objects (false negatives) or unnecessarily obscures non-identifiable objects (false positives).

False positives do not necessarily raise privacy concerns, and more so impact the quality of the product's imagery. We introduce a new failure mode that emerges from considering \textit{group privacy}, instead of only the privacy of individuals. We call this failure mode `group membership inference.' In this failure mode, a collection of de-identified objects of the same type can be clustered using computational methods, leaking the privacy guarantees of the de-identification. We offer a concrete example to illustrate this point: consider construction workers, who frequently wear high-visibility green vests. Even with the most stringent de-identification method, blurring the entire pedestrian body, a clustering algorithm could detect and group these vests. This would effectively compromise the privacy guarantees for the group of construction workers, despite individual de-identification. We summarize DSI's de-identification failure modes in \autoref{tab:obfuscation-taxonomy}.
\FloatBarrier

\section{Examples of Group-Based Privacy Violations in DSI under CI}

\noindent\fbox{%
\begin{minipage}{0.98\textwidth}
\centering
\vspace{0.5em}
\begin{tabular}{p{0.9\textwidth}}
\textbf{Protesters} \newline
\textbf{Information:} Meeting locations and times (geo-tagged). \newline
\textbf{Recipients:} Political groups opposed to the protester's cause. \newline
\textbf{Transmission Principle:} Information about meeting locations and times is shared outside the original context of trusted participants and disseminated to opposition political groups, violating expectations of confidentiality and purpose limitation. \newline
\textbf{Harms:} Retaliation, disruption of assemblies, suppression of free speech. \\ \hline
\textbf{Nurses} \newline
\textbf{Information:} Shift patterns derived from DSI near healthcare facilities. \newline
\textbf{Recipients:} Malicious employers or staffing agencies. \newline
\textbf{Transmission Principle:} Shift pattern data is aggregated and shared with malicious employers or staffing agencies without the consent of the individuals, violating principles of data minimization and appropriate recipient access. \newline
\textbf{Harms:} Exploitation of labor patterns, unsafe working conditions, reduced autonomy. \\ \hline
\textbf{Commuters} \newline
\textbf{Information:} Behavioral patterns at busy crosswalks during peak hours. \newline
\textbf{Recipients:} Predatory advertisers. \newline
\textbf{Transmission Principle:} Behavioral patterns shared with predatory advertisers violates expectations of anonymity and proportionality in the collection and use of public data. \newline
\textbf{Harms:} Exploitation through targeted marketing (e.g., payday loans, fast food). \\ \hline
\textbf{Religious groups} \newline
\textbf{Information:} Images and metadata in cultural or religious locations. \newline
\textbf{Recipients:} Hate groups. \newline
\textbf{Transmission Principle:} Sensitive imagery and metadata from religious locations shared with hate groups violate principles of contextual sensitivity, trust, and non-maleficence in handling sensitive personal data. \newline
\textbf{Harms:} Harassment, stigmatization, violence, violation of religious freedoms.
\end{tabular}
\vspace{0.5em}

\captionof{table}{Examples of Group-Based Privacy Violations in DSI under CI}
\label{tab:group-harms}
\end{minipage}
}

\end{document}